\documentclass[useAMS,usenatbib,useAMS,usegraphicx
]{mn2e}
\usepackage{amssymb}
\newcommand{\rx}{RX J0852.0-4622}             
\newcommand{\Teff}{$T_{eff}$}
\newcommand{\lgg}{log\,$g$}
\newcommand{\Vt}{$V_t$}
\newcommand{\Vsini}{$V\sin~i$}
\newcommand{\Vlsr}{v_{\rm lsr}}
\newcommand{\kms}{km~s$^{-1}$}
\newcommand{\cms}{cm~s$^{-2}$}
\newcommand{\cmt}{cm$^{-2}$}
\newcommand{\cmc}{cm$^{-3}$}
\newcommand{\header}[1]{\multicolumn{1}{c}{\textrm{#1}}}

\title[Interstellar absorptions and shocked clouds]
{Interstellar absorptions and shocked clouds towards supernova
remnant \rx \thanks{Based on observations collected at the European 
Southern Observatory, Chile, 080.D-0012(A)}
}

\author[Yu.Pakhomov et al.]{
Yu.~V.~Pakhomov,$^1$ 
N.~N.~Chugai$^1$ and
A.~F.~Iyudin$^{2,3}$
 \\
$^1$Institute of Astronomy, Russian Academy of Sciences, Pyatnitskaya 48,
119017, Moscow, Russian Federation\\
$^2$Skobeltsyn Institute of Nuclear Physics, Moscow State University, 
Vorob'evy Gory, 119992 Moscow, Russian Federation\\
$^3$Max-Planck-Institut f\"ur Extraterrestrische Physik, Postfach 1312, D-85741 
Garching, Germany
}

\begin{document}
\maketitle

\begin{abstract}
We present results of survey of interstellar absorptions towards supernova 
remnant (SNR) \rx. The distribution of K\,{\sc i} absorbers along the distance 
of the background stars is indicative of a local region ($d<600$~pc) 
strongly depopulated by K\,{\sc i} line-absorbing clouds. This fact is 
supported by the behavior of the interstellar extinction.
We find four high-velocity Ca\,{\sc ii} components with velocities of $>100$~\kms\
towards three stars and identify them with shocked clouds of Vela SNR. 
We reveal and measure acceleration of two shocked clouds at the approaching 
and receding sides 
of Vela SNR along the same sight line. The clouds acceleration, 
velocity, and Ca\,{\sc ii} column density are used to probe cloud parameters. 
The total hydrogen column density of both accelerating 
clouds is found to be similar ($\sim6\times10^{17}$~\cmt) which 
indicates that  possibly there is a significant amount of small-size 
clouds in the vicinity of Vela SNR.

\end{abstract}


\section{Introduction}

Optical absorption spectroscopy is a classical tool to probe interstellar
(IS) clouds, viz., velocities, Doppler broadening, column density of
atomic and molecular species, and temporal variations
\citep{1995AJ....109.2627D, 2003Ap&SS.285..661C}. The 
direction of Vela SNR is unique among other regions of sky in one 
important respect: apart
from usual low-velocity absorption components we see there fast-moving 
absorbers
with velocities 100-200~\kms\ \citep{1971ApJ...170..289W, 1995AJ....109.2627D,
2000ApJS..126..399C}. These components are identified with IS clouds 
shocked in
the blast wave of Vela SNR. An exciting feature of these high-velocity
components is their temporal variability \citep{1995AJ....109.2627D,
2000ApJS..126..399C}. Among these variations of particular interest is the
evidence of cloud acceleration \citep{2000ApJS..126..399C} which probably
reflects ongoing process of blast wave/cloud interaction in Vela SNR. 
Surprisingly enough, no attempts have been made so far to utilise 
acceleration of shocked clouds as an important observational constraint 
of cloud parameters. 

\begin{table*}
\begin{minipage}{180mm}
\centering
\caption{List of observed stars}
\label{tab:list}
\begin{tabular}{lccclrcccccc}
\hline
\header{Star}&    RA     &     Decl &\header{$m_v$}&
\header{SpType}&\header{$(B-V)$}& \header{$A_V$} &\header{$d$}&
\multicolumn{4}{c}{\textrm{S/N}}\\
\cline{2-3}\cline{9-12}
&\multicolumn{2}{c}{(eq. 2000.0)}&\header{(mag)}&    
&\header{(mag)}&\header{(mag)}&\header{(pc)}& all&\header{Ca\,{\sc ii}} &
\header{Na\,{\sc i}}
&
\header{K\,{\sc i}}\\
\hline
HD\,75309  &08 47 28.0&-46 27 04& 7.84&B2Ib/II &
0.01&0.8&1900$\pm$300&113&38&146&114\\
HD\,75820  &08 50 26.0&-46 14 53& 8.64&B9V     &-0.02&0.2&470$\pm$100 &79
&30&101&
78\\
HD\,75873  &08 50 48.8&-46 18 36& 8.10&A3II/III& 0.38&1.2&1400$\pm$200&67 &18&
88&
82\\
HD\,75955  &08 51 26.0&-45 37 23& 7.73&B9V     &-0.01&0.2&320$\pm$~70 &65 &20&
83&
61\\
HD\,75968  &08 51 32.8&-46 36 36& 8.14&B9III/IV&-0.12&0.0&570$\pm$140 &45 &14&
61&
57\\
HD\,76060  &08 52 02.4&-46 17 20& 7.88&B8IV/V  &-0.09&0.1&390$\pm$~90
&101&34&130&
97\\
HD\,76589  &08 55 23.0&-46 53 28& 8.34&B9IV    &-0.05&0.1&390$\pm$~90 &67 &17&
87&
66\\
HD\,76649  &08 55 50.4&-46 20 30& 8.33&B7II/III& 0.14&0.8&640$\pm$110 &57 &14&
74&
65\\
HD\,76744  &08 56 18.2&-46 19 57& 8.69&A0V     & 0.08&0.5&270$\pm$~50 &61 &18&
78&
59\\
CD-454590&08 49 35.5&-46 23 18& 9.58&B5      & 0.20&1.3&2400$\pm$300&41 &11& 53&
43\\
CD-454606&08 50 15.0&-45 31 22& 8.96&B0.5V   & 0.38&2.0&1670$\pm$160&63 &18& 82&
72\\
CD-454645&08 51 34.9&-46 09 54&10.32&A0      & 0.20&0.4&330$\pm$~70 &38 &11& 49&
41\\
CD-454676&08 53 22.0&-46 02 09& 8.93&B0.5III & 0.77&3.2&1080$\pm$150&38 &13& 52&
54\\
CD-464666&08 50 44.3&-46 38 11& 9.81&A0II    & 0.60&2.1&5700$\pm$500&37 &12& 50&
51\\
\hline                                                            
\end{tabular}
\end{minipage}
\end{table*}

Recently, we performed low- and high-resolution spectroscopic survey of 14 stars
in the field of \rx\ aka Vela Jr. The latter is superimposed on the south-east
part of the Vela SNR. The analysis of low-resolution observations in the region 
3700--4000~\AA\ was reported earlier \citep{2010A&A...519A..86I}. Here we address
the high resolution spectral data which cover much broader spectral range. 
We find
the results of analysis of these spectra are of great interest by several
reasons. Firstly, our survey covers poorly observed region of Vela SNR. Only 
two stars
HD\,75309 and HD\,75821 of the previous survey \citep{2000ApJS..126..399C} fall
into this region. Secondly, our spectra cover essentially larger spectral range
than any other previous survey in the Vela direction. This permits us 
to obtain more
extensive picture of interstellar atomic and molecular absorbers in the 
Vela SNR
direction. Thirdly, and most importantly, HD\,75309, the common star of our 
and \citet{2000ApJS..126..399C} survey, shows high-velocity Ca\,{\sc ii} 
absorption at
-120~\kms\ and +120~\kms\ presumably associated with shocked clouds. 
Our data combined with those of \citet{2000ApJS..126..399C} survey 
cover large time interval of 15~yr which open up a possibility to 
measure rather precisely the
velocity change of shocked clouds, and thus to recover the clouds 
acceleration.
In turn, the cloud acceleration value in combination with the velocity 
and Ca\,{\sc ii}
column density provides us with a new diagnostic tool for probing cloud
parameters. 

The paper is organized as follows. We first describe observations and the
extraction of interstellar lines (Section \ref{sec-obs}). In two cases 
the stars turn out slow rotators and, therefore, 
 detailed stellar atmosphere analysis 
is used to recover interstellar absorption. The parameters of interstellar
absorption components and their analysis are presented in  Section
\ref{sec-res}. In Section~\ref{sec-shock} we address the shocked clouds in the
sight line of HD\,75309. The cloud acceleration, velocity and Ca\,{\sc ii} column
density are used for probing cloud parameters on the bases of cloud shock
model in which we include all the relevant physics.

\section{Observations and data reduction}
\label{sec-obs}

The spectra of 14 stars in direction of Vela Jr. [for star position see
\citep{2010A&A...519A..86I}] were obtained in March 2008 on the 3.6-m ESO
NTT telescope equipped with EMMI spectrograph (echelle grating~\#14,
cross-dispersion grating~\#3, CCD mosaic 2076x4110). The full coverage is
3850-8620~\AA\ with the gap at 5570-5650~\AA; the central dispersion is
0.02~\AA/pix, the resolving power is $R=\lambda/\Delta\lambda$=88\,600
($\Delta\,v$=3.4~\kms). In Table~\ref{tab:list} stars are listed with their
HD/CD name, equatorial coordinates, $V$-magnitude, spectral type, $B-V$ color,
and signal-to-noise ratio (S/N), average and at Ca\,{\sc ii} 3933,3968~\AA, Na\,{\sc i}
5889,5895~\AA, and K\,{\sc i} 7664,7698~\AA\ positions. Preliminary processing of CCD
images, spectra extracting, wavelength calibration with ThAr-lamp, normalizing
and merging of orders is performed using \emph{echelle} package of \emph{MIDAS}.
In order to normalize the spectra a blaze function of the echelle is recovered
using a calibrated flux of the standard star HD\,60753. Its stellar flux is
synthesized by applying \emph{ATLAS9} code \citep{1993KurCD..13.....K} and
\emph{SynthVb} \citep{2003IAUS..210P.E49T}. A stellar atmosphere model of
standard star is calculated for the solar metallicity with parameters
\Teff=16200\,K, \lgg=3.56, \Vsini=24~\kms, and interstellar absorption
$A_V=0.25$ \citep{2010A&A...519A..86I}.

For most stars we used stellar synthetic spectrum assuming solar element
abundances because these objects belong to the thin disk of Galaxy. We find that
small deviations ($\lesssim0.3$ dex) from the solar abundance do not affect the
extracted interstellar lines, if the stellar rotation is high
($\gtrsim30$~\kms). In case of star HD\,75873 the rotation is slow, \Vsini=
3~\kms, and, therefore, a stellar synthetic spectrum with carefully determined
chemical composition should be used to extract interstellar lines. A similar
procedure is applied to CD-464666.

The stellar parameters of HD\,75873 are found by employing  a standard 
requirement
that the chemical composition should be stable against variations of the line
excitation potential and ionization state. The \emph{ATLAS9} code
is used to derive parameters of the stellar atmosphere and 
\emph{WIDTH9} \citep{1993KurCD..13.....K}
for the abundance determination. The stellar parameters infered 
in this way (\Teff=8900\,K,
\lgg=2.50) are consistent with those determined earlier using only hydrogen
lines \citep{2010A&A...519A..86I}. The microturbulent velocity (\Vt=2.5~\kms)
is derived using requirement that the Fe/H ratio should not depend on the 
intensity of Fe\,II lines. The recovered element abundance slightly exceeds the
solar value; the Fe overabundance is +0.08$\pm$0.11 dex. In case of CD-464666,
most distant star of the program, we find \Teff=10100$\pm$300\,K,
\lgg=2.0$\pm$0.2, and \Vt=2.5$\pm$0.5~\kms. The distance we infer using the
spectral parallax method is 5800$\pm$900~pc, in agreement with the previous
estimate 5700$\pm$500~pc \citep{2010A&A...519A..86I}. The inferred element 
abundance is close to the solar one; Fe overabundance is of +0.05$\pm$0.11 dex. 

\begin{table*}
\begin{minipage}{180mm}
\caption{Equivalent widths (m\AA) of interstellar atomic
and molecular lines}
\label{tab:ew_lines}
\centering
\begin{tabular}{l|cc|cc|cc|c|c}
\hline
\header{Star}&
\multicolumn{2}{|c|}{Ca\,{\sc ii}}&
\multicolumn{2}{c|}{Na\,{\sc i}}  &
\multicolumn{2}{c|}{K1}   &
CH & CH$^+$\\
\cline{2-9}\\
&
3933~\AA & 3968~\AA &
5889~\AA & 5895~\AA &
7664~\AA & 7698~\AA &
4300~\AA &
4232~\AA \\
\hline
  HD\,75309 & 297 &  131 &  567 &   450&   57 &   52 &  6 &  7 \\
  HD\,75820 &  96 &   16 &  110 &    50&  --- &  --- & -- & -- \\
  HD\,75873 & 101 &  100 &  536 &   454&  204 &  161 & 26 & 10 \\
  HD\,75955 &  87 &   21 &  199 &   116&  --- &  --- & -- & -- \\
  HD\,75968 &  62 &   17 &  122 &    74&  --- &  --- & -- & -- \\
  HD\,76060 & 108 &   22 &  192 &   110&  --- &  --- & -- & -- \\
  HD\,76589 &  85 &   28 &  115 &    63&  --- &  --- & -- & -- \\
  HD\,76649 &  41 &   65 &  383 &   283&   82 &   89 & 15 &  8 \\
  HD\,76744 &  77 &  --- &  101 &    50&  --- &  --- & -- &    \\
CD-454590 & 431 &  113 &  568 &   533&  164 &  132 & 22 & 20 \\
CD-454606 & 368 &  189 &  632 &   496&  198 &  153 & 22 & 11 \\
CD-454645 &  17 &   10 &  244 &   169&  --- &  --- & -- & -- \\
CD-454676 & 156 &  107 &  470 &   427&  228 &  208 & 37 &  6 \\
CD-464666 & 283 &  147 &  780 &   694&  177 &  136 &  6 & 17 \\
\hline                                                            
\end{tabular}
\end{minipage}
\end{table*}
\begin{table*}
\begin{minipage}{180mm}
\caption{Equivalent widths (m\AA) of DIBs}
\label{tab:ew_dib}
\centering
\begin{tabular}{l*{12}c}
\hline
\header{Star}& 
5780~\AA& 5797~\AA& 5850~\AA& 6196~\AA& 6203~\AA& 6270~\AA& 6284~\AA& 6376~\AA& 6379~\AA&
6614~\AA&6660~\AA&7224~\AA\\
\hline
HD\,75309   & 143&  20&  --&  11&  43&  26& 308&  --&  16&  52&  10&  31\\
HD\,75820   &  --&  --&  --&  --&  --&  --&  --&  --&  --&  --&  --&  --\\
HD\,75873   & 321&  75&  25&  29&  73&  62& 598&  16&  41& 118&  24&  56\\
HD\,75955   &  --&  12&  --&  --&  --&  --&  --&  --&  --&  --&  --&  --\\
HD\,75968   &  23&  --&  --&  --&  --&  --&  27&  --&  --&  --&  --&  --\\
HD\,76060   &  22&  --&  --&  --&  --&  --&  24&  --&  --&  --&  --&  --\\
HD\,76589   &  18&  --&  --&  --&  --&  --&  --&  --&  --&  --&  --&  --\\
HD\,76649   & 114&  27&  --&  --&  19&  --& 253&  --&  11&  --&  --&  --\\
HD\,76744   &  10&  --&  --&  --&  --&  --&  --&  --&  --&  --&  --&  --\\
CD-454590 & 246&  97&  18&  27&  62&  47& 620&  15&  54& 107&  30&  65\\
CD-454606 & 261&  74&  21&  24&  57&  41& 409&  18&  46& 128&  21&  44\\
CD-454645 &  --&  --&  --&  --&  --&  --&  63&  --&  --&  --&  --&  --\\
CD-454676 & 334& 134&  35&  30&  82&  37& 769&  20&  71& 151&  21&  86\\
CD-464666 & 485& 125&  30&  35& 113&  59& 966&  23&  58& 161&  33&  94\\
\hline                                                            
\end{tabular}
\end{minipage}
\end{table*}

\begin{figure*}
\centering
\resizebox{0.85\hsize}{!}{\includegraphics[clip]{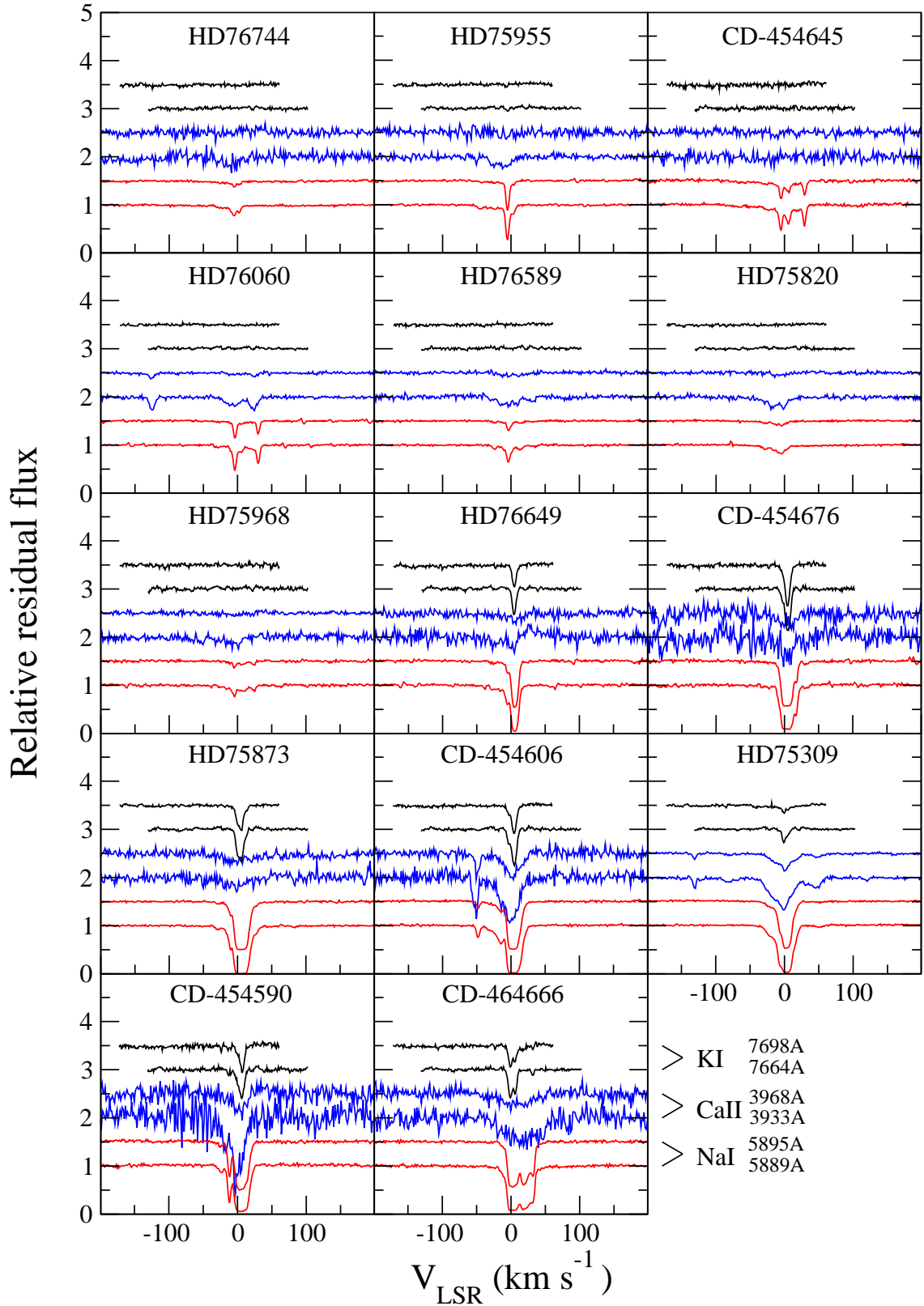}}
\caption{Interstellar absorption lines of K\,{\sc i}, Ca\,{\sc ii}, and Na\,{\sc i}
in spectra of program stars. The star distance increases from left to 
right and from top to bottom. In the right-bottom corner we show position
of spectral lines. All continuum fluxes are normalized to unity. Spectra are
shifted in vertical direction for convenience.}
\label{fig:all}
\end{figure*}

The extracted spectrum of the interstellar absorption (Fig.~\ref{fig:all})
is represented by a residual flux
$$
F(\lambda)=1-(F_{\rm syn}(\lambda)-F_{\rm obs}(\lambda))/F_{\rm
cont}(\lambda)\,, 
$$
where $F_{\rm obs}$ is the observed normalized flux of the stellar spectrum, 
$F_{\rm syn}$ is the model normalized flux calculated using \emph{SynthV} 
package and VALD line list \citep{1999A&AS..138..119K}, $F_{\rm cont}$ is the 
continuum flux. We extracted telluric lines in case of Na\,{\sc i} 5895~\AA, K\,{\sc i}
7664~\AA, and diffuse interstellar bands (DIBs).


\begin{figure}
\centering
\resizebox{\hsize}{!}{\includegraphics[clip]{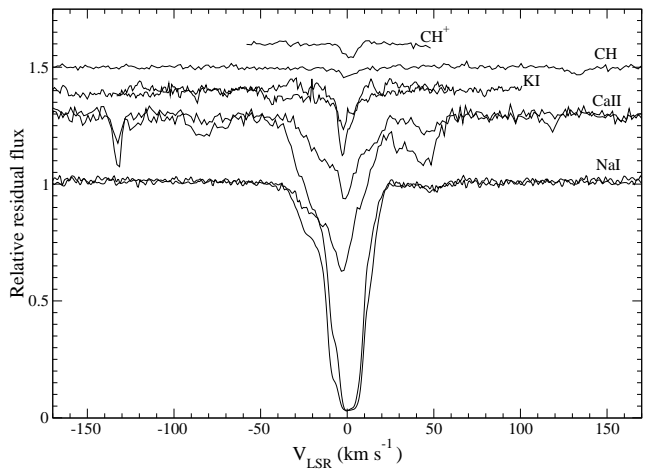}}
\caption{Interstellar absorption lines in spectra of HD\,75309. All continuum
fluxes are normalized to unity. Spectra are shifted in vertical direction for
convenience.}
\label{fig:hd75309_ism}
\end{figure}

\section{Data analysis}
\label{sec-res}

\subsection{Overview of IS absorptions}

The interstellar spectra show absorption lines of Ca\,{\sc ii} 3933, 3968~\AA, 
Na\,{\sc i} 5889, 5895~\AA, K\,{\sc i} 7664, 7698~\AA, CH 4300~\AA, CH$^+$ 4232~\AA, and DIBs.
The Ca\,I 4226~\AA\ line is not detected, in accord with the well known
weakness of this line \citep{2003ApJS..147...61W}. In Fig.~\ref{fig:hd75309_ism}
we present a zoom of  Na\,{\sc i}, Ca\,{\sc ii}, K\,{\sc i}, CH, and CH$^+$ interstellar
lines for the interesting case of HD\,75309 ($d=1900$~pc) 
 showing high-velocity
($|v|>100$~\kms) Ca\,{\sc ii} components that will be addressed below. Total
equivalent widths of the detected interstellar absorptions are given in
Table~\ref{tab:ew_lines}. 

The interstellar Na\,{\sc i} doublet is detected in the spectra of all stars
(Fig.~\ref{fig:all}). This line, as expected, becomes stronger as star 
distance
increases. In most cases Ca\,{\sc ii} doublet is significantly weaker than that of
Na\,{\sc i} which is well known fact and usually explained as an effect of a 
strong Ca depletion onto dust grains. K\,{\sc i} doublet is seen only in stars with
large extinction and strong molecular absorptions of CH and CH$^{+}$. The
equivalent width of K\,{\sc i} 7698~\AA\ absorption tightly correlates with the total
equivalent width of molecular lines of CH and CH$^+$.
(Table~\ref{tab:ew_lines}). This correlation agrees with the strong correlation
between K\,{\sc i} and CN column densities \citep{2001ApJS..133..345W}.

\begin{figure}
\centering
\resizebox{\hsize}{!}{\includegraphics[clip]{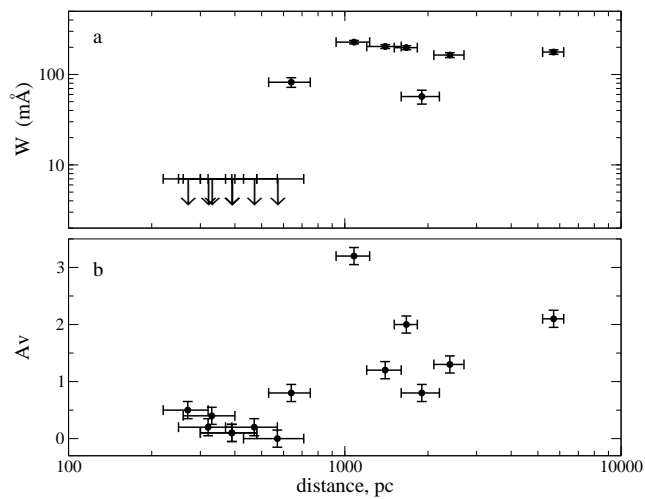}}
\caption{Equivalent width of K\,{\sc i} absorption ({\em upper panel}) and extinction
({\em lower panel}) vs. star distance. Arrows in upper panel show upper limits.}
\label{fig:KI-Av}
\end{figure}

The spectra arranged according to the star distance in Fig.~\ref{fig:all} reveal
a striking behavior of K\,{\sc i} absorptions: they are not detected 
for stars with distances
$d<600$~pc ($N({\rm K\,{\sc i}})<5\times10^{10}$~\cmt) and then sharply get
strong for stars with distances $d>600$~pc 
(Fig.~\ref{fig:KI-Av},a). 
The "jump" in the distribution of K\,{\sc i} absorbers along the distance 
is indicative of a "cavity" in the region $d<600$~pc underpopulated by 
cool clouds. This conjecture is supported by a similar behavior  
of the interstellar extinction although in this case the threshold effect 
is not as strong as for K\,{\sc i} absorbers (Fig.~\ref{fig:KI-Av},b). 

Twelve diffuse interstellar bands are detected in our sample and listed in
Table~\ref{tab:ew_dib}. Empty place in the table indicates that this 
particular 
DIB is absent or its equivalent width is smaller than 10~m\AA. Most DIBs with
$\lambda<$5500~\AA\ cannot be extracted because of low S/N ratio. In an 
extensive survey of DIBs in spectra of $\sim200$ stars
\citep{2011ApJ...727...33F} finds a strong correlation of equivalent widths 
($W$ in m\AA) of 6284~\AA\ and 5780~\AA\ DIBs
(correlation coefficient $r=0.96$) with the linear fit
$W(6284)=a+bW(5780)$, where $a=28.24\pm5.8$ and $b=2.32\pm0.03$. In our small
sample the correlation even stronger ($r=0.99$), while the parameters of the
linear fit are a bit different: $a=1.5\pm46$ and $b=2.03\pm0.18$. Yet the
relatively small value of $a$ term in our case seems to be more easy to conform
with the expected zero value of $W$ in case of vanishing molecular abundance.

\subsection{Component parameters}

\citet{2001MNRAS.328.1115C} demonstrated upon the bases of high-resolution
($\Delta v\approx$0.3~\kms) observations that Doppler parameters of interstellar
line components lie in the range of $0.3\leq b\leq 2.2$~\kms. Our spectra 
have lower resolution ($\Delta v=3.4$~\kms) so most of the interstellar line
components are expected to be unresolved. This precludes reliable determination
of the Doppler parameter of a single line. Fortunately, in cases of Na\,{\sc i},
Ca\,{\sc ii}, and K\,{\sc i} we have in hand a doublet ratio; this provides us with the
additional relation which permits one to recover both $N$ and $b$ even for 
unresolved lines. Generally, the blend of $M$ components is decomposed using 
standard relations
\begin{eqnarray}
\label{synth_F}
&F(v) &= (\mbox{e}^{\displaystyle -\tau(v)}\ast\Theta) \\
\label{synth_tau}
&\tau(v)&=\sigma_0f\lambda\sum_{k=1}^{M}N_k\phi\left(\frac{v-v_k}
{b_k}\right)\,,
\end{eqnarray}
where the first equation is a convolution of the line profile with the
instrumental profile $\Theta$. In the second equation $\sigma_0=0.0265$ is the 
integrated cross-section parameter, $f$ is the oscillator strength taken from
VALD database \citep{1999A&AS..138..119K} for atomic lines and from
\citet{1993A&A...269..477G} for molecular lines, $N_k$ is the column density  of
$k$-th component, and $\phi(u/b)=(1/b\sqrt\pi)\exp(-u^2/b^2)$ is the Gaussian
velocity distribution of an absorbing gas in a cloud along the sight line.

Parameters of interstellar absorption components derived by the 
minimization procedure are listed in
Table~\ref{tab:components}. The radial velocity is reduced to the local standard
of rest (LSR) using the Sun velocity towards Vela Jr. $\Delta\Vlsr=-15.7$~\kms\
\citep{2005A&A...430..165F}. For each atomic species three columns 
($\Vlsr$, $b$, and  $N$) are given. The Doppler parameter for molecular lines is omitted
because it cannot be reliably determined. Data with the large errors 
are marked by the colon. The typical $\Vlsr$ error is 0.1--0.2~\kms\
(0.5--1~\kms\ in case of the low accuracy), $b$ error is 0.1--0.4~\kms\
(1--1.5~\kms\ in case of the low accuracy);  for $N$ a typical relative error is
10\%  and 30\% in case of the low accuracy. The blends with unresolved saturated
lines are marked by italic with the estimated velocities shown in parentheses.

\begin{table*}
\renewcommand{\tabcolsep}{4pt}
\renewcommand{\arraystretch}{0.9}
\caption{Parameters of the resolved components of interstellar
absorptions}
\label{tab:components}
\begin{tabular}{c|*{3}{c|}|*{3}{c|}|*{3}{c|}|*{2}{c|}|*{2}{c}}
\hline
Star   & \multicolumn{3}{c|}{Ca\,{\sc ii}}  & \multicolumn{3}{c|}{Na\,{\sc i}}  &
\multicolumn{3}{c|}{K\,{\sc i}}  & \multicolumn{2}{c|}{CH}  & \multicolumn{2}{c|}{CH+}
\\
\cline{2-4}\cline{8-10}\cline{13-14}
 & $\Vlsr$ & $b$ & $N_{11}$  & $\Vlsr$ & $b$ & $N_{11}$  & $\Vlsr$ & $b$ &
$N_{11}$  & $\Vlsr$ & $N_{11}$  & $\Vlsr$ & $N_{11}$ \\
 &  {\tiny(\kms)}    &{\tiny(\kms)}& {\tiny(\cmt)}  &  {\tiny(\kms)}   
&{\tiny(\kms)}& {\tiny(\cmt)}&{\tiny(\kms)} &{\tiny(\kms)}& {\tiny(\cmt)}
& {\tiny(\kms)}& {\tiny(\cmt)} & {\tiny(\kms)} & {\tiny(\cmt)} \\
\hline
   HD\,75309 &  -127.1 &  1.6 &   2.0 &    -17: &      &       &     2.7 &  1.5
& 
 1.5 &    5.0: &   75: &     6.6 &    96 \\
           &    -77: &   2: &    1: &    -2.6 &      &       &     8.6 &  2.0 & 
 0.6 &         &       &         &       \\
           &    -65: & 1.5: &  0.5: & \multicolumn{3}{l}{\it2...11(5.0:~8.8:)}
& 
       &      &       &         &       &         &       \\
           & \multicolumn{3}{l}{\it-28...25(-22:~-9:~2.5:~12.5:)} &     20: &  
3: &    3: &         &      &       &         &       &         &       \\
           & \multicolumn{3}{l}{\it33...55} &         &      &       &        
& 
    &       &         &       &         &       \\
           &   123.9 & 1.5: &   0.6 &         &      &       &         &      & 
     &         &       &         &       \\
\hline
   HD\,75820 &   -14.9 &      &  3.0: &  -27.1: & 1.0: &  0.5: &         &     
& 
     &         &       &         &       \\
           &     2.0 &      &   3.0 &  -22.2: & 1.0: &  0.5: &         &      & 
     &         &       &         &       \\
           &         &      &       &  -13.6: &  1.5 &  0.6: &         &      & 
     &         &       &         &       \\
           &         &      &       &   -7.6: & 1.0: &  0.9: &         &      & 
     &         &       &         &       \\
           &         &      &       &    -1.0 &  1.1 &  1.1: &         &      & 
     &         &       &         &       \\
           &         &      &       &     4.5 & 1.0: &  0.5: &         &      & 
     &         &       &         &       \\
           &         &      &       &     8.2 & 1.0: &  0.4: &         &      & 
     &         &       &         &       \\
\hline
   HD\,75873 &    1.4: &      &   13: &   -25.6 & 3.0: &   0.6 &     3.2 &  1.5
& 
 4.4 &    4.5: &  400: &     6.1 &   158 \\
           &         &      &       &    -6.8 & 2.3: &  4.8: &     9.1 &  1.5 & 
 7.6 &         &       &         &       \\
           &         &      &       & \multicolumn{3}{l}{\it-3...25} &    15.4 &
0.7: &   1.1 &         &       &         &       \\
           &         &      &       &    30.2 & 3.0: &  1.5: &         &      & 
     &         &       &         &       \\
\hline
   HD\,75955 &    -21: &      &   3.5 &   -39.2 &  1.4 &  0.4: &         &     
& 
     &         &       &         &       \\
           &     -3: &      &   5.5 &   -19.3 &  1.3 &  0.5: &         &      & 
     &         &       &         &       \\
           &         &      &       &    -1.1 &  1.4 &  16.5 &         &      & 
     &         &       &         &       \\
           &         &      &       &     7.5 &  1.7 &   1.2 &         &      & 
     &         &       &         &       \\
\hline
   HD\,75968 &    1.0: &      &  6.0: &   -12.8 & 1.0: &   0.4 &         &     
& 
     &         &       &         &       \\
           &         &      &       &    -6.7 & 1.5: &   0.4 &         &      & 
     &         &       &         &       \\
           &         &      &       &    -1.6 &  1.0 &   1.7 &         &      & 
     &         &       &         &       \\
           &         &      &       &    10.8 & 1.5: &   0.7 &         &      & 
     &         &       &         &       \\
           &         &      &       &    27.2 & 0.5: &   0.8 &         &      & 
     &         &       &         &       \\
\hline
   HD\,76060 &  -121.8 &   4: &    4: &    -0.8 &  1.5 &   5.6 &         &     
& 
     &         &       &         &       \\
           & \multicolumn{3}{l}{\it-15...7(-8:~-1:)} &     9.1 & 2.5: &  1.0: &
 
      &      &       &         &       &         &       \\
           &   26.4: &   7: &  6.3: &    33.2 &  0.9 &   3.9 &         &      & 
     &         &       &         &       \\
\hline
   HD\,76589 & \multicolumn{2}{l}{\it-27...13} &   -21.1 &  2.0 &   0.5 &      
 &
     &       &         &       &         &       \\
           &     35: &      &  1.4: &    -7.5 &  1.5 &  0.5: &         &      & 
     &         &       &         &       \\
           &         &      &       &    -1.2 &  1.4 &   2.6 &         &      & 
     &         &       &         &       \\
           &         &      &       &     3.8 &  0.7 &  0.8: &         &      & 
     &         &       &         &       \\
           &         &      &       &    16.1 &  3.5 &   0.7 &         &      & 
     &         &       &         &       \\
\hline
   HD\,76649 &         &      &       &    -2.2 &  1.3 &   3.8 &     6.9 & 1.0:
& 
6.7: &     8.1 &   220 &     8.0 &   130 \\
           &         &      &       & \multicolumn{3}{l}{\it1...16} &    10.6 &
0.3: &  1.5: &         &       &         &       \\
\hline
   HD\,76744 &   -5.0: &      &       & \multicolumn{2}{l}{\it-22...-15} &      
 
&      &       &         &       &         &       \\
           &         &      &       &    -6.5 &  1.9 &   0.6 &         &      & 
     &         &       &         &       \\
           &         &      &       &    -1.7 &  1.5 &   1.2 &         &      & 
     &         &       &         &       \\
           &         &      &       &     5.2 &  2.6 &   1.0 &         &      & 
     &         &       &         &       \\
\hline
 CD-454590 & \multicolumn{3}{l}{\it-15...17(-13.5:~2:~9:)} &   -20.2 &  2.5 &  
1.2 &    -8.0 &  0.8 &  0.7: &    3.1: &   85: &     8.0 &   290 \\
           &         &      &       &    -7.8 &  1.3 &    50 &    -1.4 & 0.7: & 
0.7: &    11.2 &   280 &         &       \\
           &         &      &       & \multicolumn{3}{l}{\it-2...27} &     4.6 &
0.8: &  0.7: &         &       &         &       \\
           &         &      &       &         &      &       &     6.7 & 0.6: & 
1.1: &         &       &         &       \\
           &         &      &       &         &      &       &    10.5 & 0.6: & 
 29: &         &       &         &       \\
           &         &      &       &         &      &       &    14.4 & 0.6: & 
 1.3 &         &       &         &       \\
\hline
 CD-454606 &  -46.2: &  3.5 &   19: &   -43.6 &  1.0 &   1.8 &     1.6 &  2.2 & 
 1.9 &     3.4 &   111 &     6.6 &   120 \\
           & \multicolumn{3}{l}{\it-14...25} &   -10.1 &  1.0 &   3.3 &     9.2
&
 2.5 &   9.1 &     9.5 &   223 &         &       \\
           &         &      &       & \multicolumn{3}{l}{\it-3...28} &        
& 
    &       &         &       &         &       \\
\hline
 CD-454645 &         &      &       &    -1.8 &  1.2 &   5.5 &         &      & 
     &         &       &         &       \\
           &         &      &       &     8.9 &  3.0 &   3.4 &         &      & 
     &         &       &         &       \\
           &         &      &       &    32.6 &  1.1 &   3.9 &         &      & 
     &         &       &         &       \\
\hline
 CD-454676 & -177.3: &      &    8: & \multicolumn{3}{l}{\it-8...16} &
\multicolumn{3}{l}{\it0...12(0.4:~1:~7.5)} &     7.2 &   580 &     8.8 &   135
\\
           &      6: &      &       &    20.1 &  1.6 &   7.9 &         &      & 
     &         &       &         &       \\
           &         &      &       & \multicolumn{3}{l}{\it33...38} &        
& 
    &       &         &       &         &       \\
\hline
 CD-464666 & \multicolumn{3}{l}{\it-25...50} &
\multicolumn{3}{l}{\it-3...38(1:~21:~35:)} &     2.2 &  1.3 &   6.0 &     4.8 &
 
    &     6.7 &       \\
           &         &      &       &         &      &       &     9.2 &  2.2 & 
 3.7 &     32: &       &         &       \\
           &         &      &       &         &      &       &    34.7 &  0.7 & 
 0.6 &         &       &         &       \\
\hline
\end{tabular}
\end{table*}

\begin{figure}
\centering
\resizebox{\hsize}{!}{\includegraphics[clip]{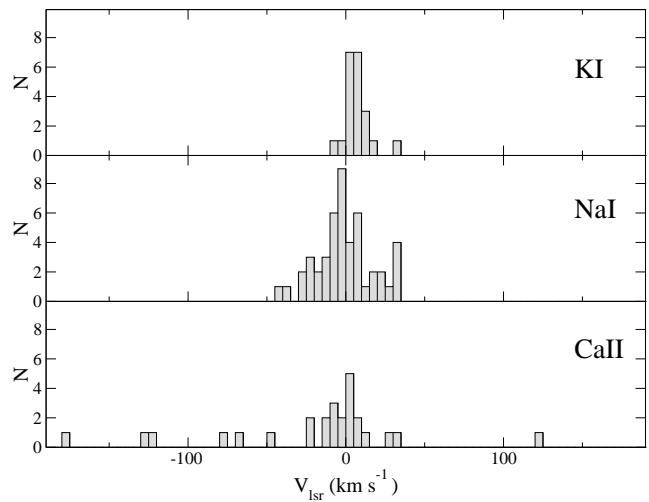}}
\caption{Velocity distribution of resolved components of interstellar
absorptions of K\,{\sc i}, Na\,{\sc i}, and Ca\,{\sc ii}.}
\label{fig:Vlsr}
\end{figure}

The average Na\,{\sc i} column density of components in our sample
($3.5\times10^{11}$~\cmt) is larger compared to $1.2\times10^{11}$~\cmt, the
average value for the high-resolution survey of \citet{1994ApJ...436..152W}. 
We attribute this disparity to the low resolution of our spectra which 
leads to the omission of the significant number of thin cloud components 
in absorption
blends. This effect is apparent also in the velocity distribution
(Fig.~\ref{fig:Vlsr}). K\,{\sc i} absorptions are rather weak so the components are
recovered confidently. The velocity dispersion of K\,{\sc i} absorbers 
$\sigma_v\approx8$~\kms\ is the same as the value 7.8~\kms\ found 
for an extensive survey for K\,{\sc i} components with $\log{N}<11.3$~\cmt\ 
\citep{2001ApJS..133..345W}. For stronger components these authors find somewhat
lower value $\sigma_v=5.8$~\kms. Our small sample precludes more refined
analysis of the velocity distribution of strong K\,{\sc i} components. In case of
Na\,{\sc i} the velocity distribution (Fig.~\ref{fig:Vlsr}) is rather broad with
$\sigma_v\approx18$~\kms, twice as large compared to 
$\sigma_v\approx9$~\kms\ found by \citet{1994ApJ...436..152W} in an extensive
survey of the Na\,{\sc i} IS absorption. The striking disparity is caused probably 
by the relatively low resolution and S/N of our spectra which precludes
decomposition of saturated blends of Na\,{\sc i} absorptions. As a result we strongly
underestimate the number of low-velocity components of Na\,{\sc i}.

The large velocity dispersion of Ca\,{\sc ii} absorption components
(Fig.~\ref{fig:Vlsr}) stems from two effects. The first is the same selection 
effect as in case of Na\,{\sc i} absorption, while the second effect is the well 
known phenomenon discussed 
\citet{1952ApJ...115..227R}: the larger ratio 
$N({\rm Ca\,{\sc ii}})/N({\rm Na\,{\sc i}})$ 
for components with higher velocity. A standard explanation of the 
Routly-Spitzer effect is the
dust destruction in the dynamically disturbed clouds which results in release 
of gaseous Ca. The fact that we see
high-velocity components in Ca\,{\sc ii} lines and not in Na\,{\sc i} lines 
indicates that
high-velocity clouds have low neutral hydrogen column density in which case the
equivalent width of Na\,{\sc i} absorption turns out below the detection 
limit of our data.  

In some cases positive velocities of $\Vlsr<50$~\kms\ are related with a
velocity gradient caused by the galactic rotation. For the most distant star
CD-464666 ($d=5700$~pc) adopting galactocentric distance of 8.5~kpc and Galaxy
rotational velocity of 220~\kms\ we find that Galactic rotation can be responsible
for the positive LSR velocities in the range of $v\leq50\pm4$~\kms. Remarkably,
this value is consistent with the upper limit of the observed velocity interval
of the Ca\,{\sc ii} absorption (-25...+50~\kms) towards CD-464666.

Two nearest stars of our sample, CD-454645 ($d=330\pm70$~pc) and HD\,76060
($d=335\pm63$~pc), show a strong Na\,{\sc i} component with similar velocity
(32.6~\kms\ and 33.2~\kms), Doppler parameter (1.1~\kms\ and 0.9~\kms), and
column density ($3.9\times10^{11}$~\cmt). This suggests that we see both stars
through the same IS cloud. For the distance of $\approx$330~pc the angular
separation $9'$ between these stars corresponds to the linear scale of 0.85~pc.
We conclude therefore that the tangential size of this cloud $\gtrsim 1$~pc. The
large LSR velocity (33~\kms) indicates that this cloud belongs to the
dynamically disturbed ISM. Remarkably, however, that we do not detect 
similar component in Ca\,{\sc ii} suggested by Routly-Spitzer effect. This
indicated that this fast-moving material contains significant amount 
dust with the most Ca locked in grains.

Of particular interest are four fast-moving Ca\,{\sc ii} components
(Table~\ref{tab:components}) with velocities exceeding 100~\kms\ in sight lines
of three stars, specifically, -177~\kms\ (CD-454676), -125 and +120~\kms\
(HD\,75309), and -122~\kms\ (HD\,76060). Following the earlier conjecture invoked
for similar high-velocity components \citep{1971ApJ...170..289W,
1976ApJ...209L..87J, 2000ApJS..126..399C} towards Vela SNR we attribute the
fast-moving material to the IS clouds shocked by Vela SNR.

Alternatively, high velocity clouds of our survey might be attributed to the 
SNR \rx. We consider, however, this possibility unlikely because \rx\
is a young SNR
that expands with the velocity of $\sim10^4$~\kms\ \citep{1998Natur.396..142I,
1998Natur.396..141A}. In this case velocities of shocked clouds unlikely 
would be the
same as in old Vela SNR. The identification of detected high-velocity components
with Vela SNR is thus preferred.

\subsection{Acceleration of high-velocity clouds in sight line of HD\,75309}

\begin{table}
\caption{Evolution of high-velocity components in spectra of HD\,75309}
\label{tab:evol}
\tabcolsep=1mm
\centering
\begin{tabular}{c|c|c|c||c|c|c}
\hline
\header{Year} & \multicolumn{3}{c}{cloud~A} & \multicolumn{3}{c}{cloud~B} \\
& \header{$\Vlsr$} & \header{$W$} & \header{$N$} &
\header{$\Vlsr$} & \header{$W$} & \header{$N$}\\
& (\kms) & (m\AA) & ($10^{11}$\cmt) & (\kms) & (m\AA) & ($10^{11}$\cmt) \\
\hline
1993 & -121.9$\pm$0.3  & 12$\pm$1 & 1.4$\pm$0.1 & 119.3$\pm$0.3  & 9$\pm$1 &
1.0$\pm$0.1\\
1996 & -123.8$\pm$0.1  & 16$\pm$1 & 1.8$\pm$0.1 & 119.9$\pm$0.3  & 7$\pm$1 &
0.8$\pm$0.1\\
2008 & -127.1$\pm$0.1  & 16$\pm$1 & 1.8$\pm$0.1 & 123.9$\pm$0.1  & 5$\pm$1 &
0.6$\pm$0.1\\
\hline
\end{tabular}
\end{table}

\begin{figure}
\centering
\resizebox{\hsize}{!}{\includegraphics[clip]{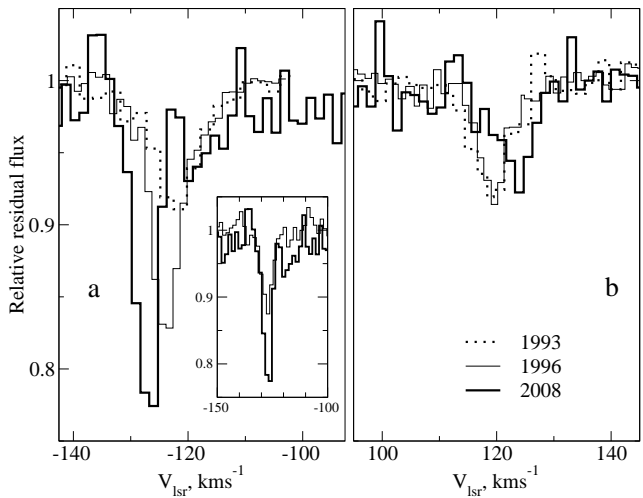}}
\caption{The profiles of Ca\,{\sc ii} absorption in spectra of star HD\,75309
for three epochs. For all the epochs Ca\,{\sc ii}~K are shown. In the inset
we show both K ({\em thick}) and H Ca\,{\sc ii} lines in 2008 spectrum to
demonstrate that the noise does not affect the narrow component position.}
\label{fig:hd75309_move}
\end{figure}

Interstellar lines towards HD\,75309 were observed in February 1993 and 
January/February 1996 \citep{2000ApJS..126..399C} with the spectral resolution 
comparable with that of our data. Authors noticed increase of a column density of
Ca\,{\sc ii} fast component with velocity -120~\kms\ but did not mention any velocity
variation. We combine their data with our observations to look at the velocity
change on the time scale of 12--15~yr. It should be noted that velocities
reported by \citet{2000ApJS..126..399C} are larger compared to ours by 4~\kms\
for all the components. The major part of this difference (2.6~\kms) is
explained by the new value we use for the solar motion
\citep{2005A&A...430..165F}.
The remaining 1.4~\kms\ which is comparable with one pixel is unclear but could
be related with the wavelength calibration uncertainty of both spectra.  Despite
this uncertainty the velocity change of the fast components can be measured with
high precision via the cross-correlation technique because slow components of
interstellar absorption lines remain unchanged. This procedure permits us to
establish a common zero point in the velocity space and thus to measure the
velocity change over the past period of time.

The high-velocity components of -127~\kms\ and +124~\kms\ according to 
our data and to those of \citet{2000ApJS..126..399C} with the common zero point
are shown in Fig~\ref{fig:hd75309_move}. We name the approaching component
"cloud~A" and receding component "cloud~B". The plots clearly demonstrate that
both components accelerate. Remarkably, in the case of cloud~A both Ca\,{\sc ii} K and
H lines in the 2008 spectrum show similar narrow component thus strengthening
the reliability of the velocity determination. Using Gaussian fitting we measure
velocity of both clouds at three epochs (Table~\ref{tab:evol}). In the table we
give also $W$ and $N$; the data on $W$ and $N$ for epochs 1993 and 1995 are
taken from \citet{2000ApJS..126..399C}. The average acceleration between 1993
and 2008 is $-0.33\pm$0.06~\kms\,yr$^{-1}$ for cloud~A and
$+0.38\pm$0.06~\kms\,yr$^{-1}$ for cloud~B. The approaching component seems to
change its width while  the equivalent width is preserved. The equivalent width
of the receding component has noticeably decreased. 

The impact parameter (in SNR radius) of the sight line of HD\,75309 relative 
to the Vela SNR
center is $p/R\sim0.5$. Adopting $p/R=0.5$ and assuming the model
of a plane slab perpendicular to the SNR radius  
we find for the cloud~A the
deprojected velocity $v_{\rm c}=147$~\kms,  Ca\,{\sc ii} column density 
$N(\mbox{Ca\,{\sc ii}})=1.6\times10^{11}$~\cmt, and the cloud acceleration
$g$=0.38~\kms\,yr$^{-1}$ (or $1.2\times10^{-3}$~\cms). For the cloud~B these
values are $v_{\rm c}=142$~\kms, $N(\mbox{Ca\,{\sc ii}})=6.9\times10^{10}$~\cmt, and 
$g=0.44$~\kms\,yr$^{-1}$ (or $1.4\times10^{-3}$~\cms). The deprojected 
values will be used below for the constraining the cloud parameters.

\section{Parameters of shocked clouds}
\label{sec-shock}

The question we address here is what are the cloud parameters (density and size)
which might explain simultaneously the Ca\,{\sc ii} column density and  the
acceleration of shocked clouds. Below we rely on the assumption that the
interstellar clouds are shocked and accelerated by the blast wave of Vela SNR.
We start with the general picture and then use a model of the cloud shock to
constrain the cloud parameters.

\subsection{General picture}

We follow a view that the Vela SNR is the result of an expansion of the Sedov
blast wave in the cloudy IS medium \citep{1999A&A...342..839B}. 
 In our estimates and modelling we approximate sub-parsec 
clouds by a homogeneous 
sphere which is the idealization for the clumpy cloud. 
The assumption of a spherical cloud is usually adopted for the 
analysis of a blast wave/cloud interaction \citep{1994ApJ...420..213K}, 
particularly in case of a shocked cloud in the Cygnus Loop 
\citep{2005ApJ...633..240P}. In fact, since we are interested primarily in the 
cloud size along the blast wave propagation we do not concern about the
clump shape and tangential size. Some justification for the choice of clumpy 
shape provide 3-dimensional 
magneto-hydrodynamic simulations \citep{2011IAUS..270..179K} which
demonstrate that initially homogeneous medium 
evolves towards a two-phase structure in which cool phase grows 
filamentary but later on the clumpy structure gets dominant.
Yet we admit that in some cases a cloud shape may 
be close to that of a filament or sheet.

A blast wave
propagating with the speed $v_{\rm b}$ in an intercloud medium with the
density $\rho_{\rm i0}$ drives a cloud shock wave with the speed $v_{\rm
s}\approx v_{\rm b}(\rho_{\rm i0}/\rho_{\rm c0})^{1/2}$
\citep{1994ApJ...420..213K} where $\rho_{\rm c0}$ is the undisturbed cloud
density. (Note, we distinguish between observed cloud velocity $v_{\rm c}$ and
the cloud shock velocity $v_{\rm s}$). Adopting the blast wave speed $v_{\rm
b}=600$~\kms\ \citep{1999A&A...342..839B} and the cloud shock velocity $v_{\rm
s}=140$~\kms\ one finds from the momentum conservation that the 
cloud-to-intercloud density contrast $\chi\equiv\rho_{\rm c0}/\rho_{\rm
i0}\approx (v_{\rm b}/v_{\rm s})^2 \approx18$. For a typical intercloud density
$n_{\rm i0}=0.1$ \citep{1999A&A...342..839B} the cloud density is then $n_{\rm
c0}\sim2$~\cmc.

The interaction of the blast wave with the cloud of the radius $a$ can be  
divided roughly into three phases \citep{1994ApJ...420..213K}. At the first
stage the cloud shock propagates through the cloud; this stage ends up with the
cloud crush in a "cloud crushing time" $t_{\rm cc}=a/v_{\rm s}$. At the second
stage the shocked cloud is accelerated by the dynamical pressure of the blast
wave. This is accompanied by the growth of the Kelvin-Helmholtz instability that
leads to the cloud fragmentation. The third stage is the final stage of 
fragmentation which ends up with the full destruction of a cloud and mixing of 
cloud fragments with the blast wave flow. 
For the density contrast $10\leq\chi\leq 100$
the cloud life time is $t_{\rm d}\sim4t_{\rm cc}$ \citep{1994ApJ...420..213K}.
It should be noted that the shocked cloud may not show acceleration. 
This would be the case if the time $t_{\rm cc}$ is so large that 
the cloud has enough time to cross the blast wave before it gets fragmented. 
The fact that we see the accelerating Ca\,{\sc ii} line-absorbing gas indicates that
the cloud has been already completely shocked and still resides inside 
the dense part of the blast wave. We however are not able to
exactly specify the stage at which the shocked cloud is detected. The
conservative statement would be that the accelerating cloud is caught at the
stage $t_{\rm cc}<t<4t_{\rm cc}$ after the interaction between the blast wave
and the cloud has turned on.

The cloud acceleration combined with an additional information about 
Vela SNR can provide us with an estimate
of the cloud column density via the equation of a cloud motion driven by the
blast wave. For the blast wave postshock velocity of $(3/4)v_{\rm b}$ and the
postshock density of $4n_{\rm i0}$ the equation of the cloud motion reads 
\begin{equation}
gN_{\rm H}=4n_{\rm i0}\left(\frac{3}{4}v_{\rm b}-v_{\rm c}\right)^2\,,
\label{eq-accel}
\end{equation}
where $N_{\rm H}$ is the cloud total hydrogen column density normal to the shock
plane. Inserting into this equation $g\approx10^{-3}$~\cms, $v_{\rm
c}=140$~\kms, $n_{\rm i0}=0.1$~\cmc, $v_{\rm b}=600$~\kms\ one infers 
the cloud column density $N_{\rm H}\approx5\times10^{17}$~\cmt. This 
exciting result shows that the accelerated cloud is unlike 
a typical Na\,{\sc i} absorber which is associated with a cloud column density
$N_{\rm H}\sim 8\times10^{19}$~\cmt\ \citep{1994ApJ...436..152W}. 
The derived estimate of $N_{\rm H}$ combined with the cloud density $n_{\rm
c0}\sim2$~\cmc\ implies the cloud radius $a=(3/4)N_{\rm
H}/n_{c0}\sim2\times10^{17}$~cm assuming initial spherical cloud
shape. For this radius the cloud crushing time is $t_{\rm cc}=a/v_{\rm c}\sim
500$~yr.

The above estimates presume a constant dynamical pressure $p_{\rm dyn}=\rho v^2$
of the blast wave. This assumption, however, could easily break down because the
dynamical pressure in the blast wave drops downstream rather steeply,  $p_{\rm
dyn}\propto(r/R)^{11}$ \citep{1987ApJ...318..674M}. Already at $r=0.97R$ the
pressure is $\approx30$\% lower compared to the maximum value at the shock
front. It is reasonable to adopt the postshock layer $\Delta\,r\sim0.03R$ to be
the width of the blast wave in which the dynamical pressure is approximately 
constant. Because $v_{\rm c}\ll v_{\rm b}$, it takes
$\Delta\,t\sim\Delta\,r/v_{\rm b}\sim$800~yr for the cloud in the blast wave
frame to cross the blast wave width $\Delta\,r=0.03R\approx0.5$~pc, where we
adopt $R=16$~pc \citep{1999ApJ...515L..25C}. The found time scale is only
slightly larger than the estimated cloud crushing time $t_{\rm cc}\sim 500$~yr.
This has an important implication for the cloud sizes: the initial radius of
accelerating clouds cannot significantly exceed $2\times10^{17}$~cm; otherwise
the cloud crosses the blast wave before it gets shocked. In this regard it
might well be that not all the high-velocity shocked clouds accelerate; 
some of them 
might be large enough to cross the blast wave before they get fragmented. 
These clouds obviously should not experience noticeable acceleration.

\subsection{Model outline}

The cloud postshock flow will be modelled in a steady plane shock approximation. 
The major
parameters of the shock are the observed cloud shock velocity ($v_{\rm s}$),
preshock cloud density ($n_{\rm c0}$), preshock magnetic field ($B_0$), 
and preshock
hydrogen ionization fraction ($x_0$). The postshock flow then is determined by a
standard set of equations of mass, momentum and energy conservation
\citep{1972ApJ...178..143C}. We adopt the parallel frozen-in
magnetic field. Cooling function of \citet{1993ApJS...88..253S} and the 
low-temperature ($\log{T}<3.9$~K) cooling function of \citet{1972ApJ...174..365J}
are used for the cooling rate calculation. We ignore effects of elements
depletion onto dust in the cooling function because in the temperature range of
interest the cooling function is determined by elements (H, He, C, N, O) which
are not affected by the depletion. The hydrogen, helium, Ca, and Na ionization
fractions are calculated by solving time-dependent kinetic equations which
include radiative recombination, ionization by electron collisions, ionization
by the interstellar ultraviolet radiation, by cosmic rays and background X-rays.
Among the latter two mechanisms the ionization by background X-rays dominates
for small size clouds ($\leq10^{18}$~cm). We adopt the ionization rate by
background X-rays to be $\zeta\approx7\times10^{-16}$~s$^{-1}$
\citep{1995ApJ...443..152W}. The ionization of Ca\,{\sc ii} by the recombination Lyc
radiation is also taken into account although this process is of minor
importance. Four Ca ions and three Na ions are included in the ionization
calculations of these elements. 

The interstellar calcium is usually strongly depleted onto dust. We adopt
depletion factors $\delta=\mbox{(Ca/H)/(Ca/H)}_{\odot}=2\times10^{-4}$ for Ca
and $\delta=0.1$ for Na \citep{1996ApJ...470..893S}. The dust destruction in the
shock is treated following standard recipes \citep{1994ApJ...431..321T}. We take
into account the grain acceleration by the betatron mechanism
\citep{1977ApJ...215..805S} and the grain deceleration due to the gas-grain 
collision. The efficiency of the betatron acceleration is set to be unity, if
the grain Larmor radius is small, $r_B(d\ln\,B/dz)<1$, and zero otherwise. The
major mechanisms of a dust destruction are the grain evaporation in grain-grain
collisions, nonthermal grain sputtering by collisions with He, and the thermal
dust sputtering \citep{1979ApJ...231...77D, 1994ApJ...431..321T}. The adopted
critical velocity of the vaporization in the grain-grain collision is
$v_v=30$~\kms\ \citep{1995ApJ...454..254B}. The surface binding energy relevant
to the grain sputtering is set to be 4~eV, the value typical for amorphous
carbon and iron grains \citep{1994ApJ...431..321T}. Calcium locked in the dust is
assumed to be homogeneously mixed with the grain material. 

An approximation of a single grain size with the initial grain radius of
$10^{-5}$~cm is used. The preshock hydrogen ionization fraction is set to 
$x_0=0.1$ and the magnetic field is set to $B_0=3\times10^{-6}$~G; in fact, 
results are not very much sensitive to these parameters. In order to estimate
the cloud acceleration we adopt the dynamical pressure of the blast wave to be
$\rho_{\rm c}v_{\rm s}^2$ and the intercloud blast wave speed of 600~\kms.

\subsection{Results of modelling}

\begin{figure}
\centering
\resizebox{\hsize}{!}{\includegraphics{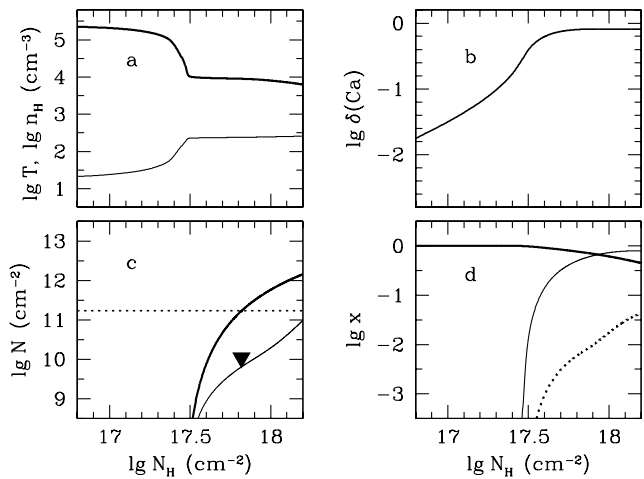}}
\caption{The postshock flow for the shock speed of 140~\kms\ as
a function of the downstream column density. Shown are
({\bf a}) temperature ({\em thick line})
and density; ({\bf b}) Ca depletion factor; ({\bf c}) Ca\,{\sc ii} column density
({\em thick solid}), Na\,{\sc i} column density ({\em thin solid}), observed Ca\,{\sc ii}
column density ({\em dotted}), and Na\,{\sc i} column density $3\sigma$ upper limit
({\em triangle}); ({\em d}) ionization fraction of H ({\em thick solid}),
Ca\,{\sc ii} ({\em thin solid}), and Ca\,I ({\em dotted}).}
\label{fig:fsh1}
\end{figure}
\begin{figure}
\centering
\resizebox{\hsize}{!}{\includegraphics[]{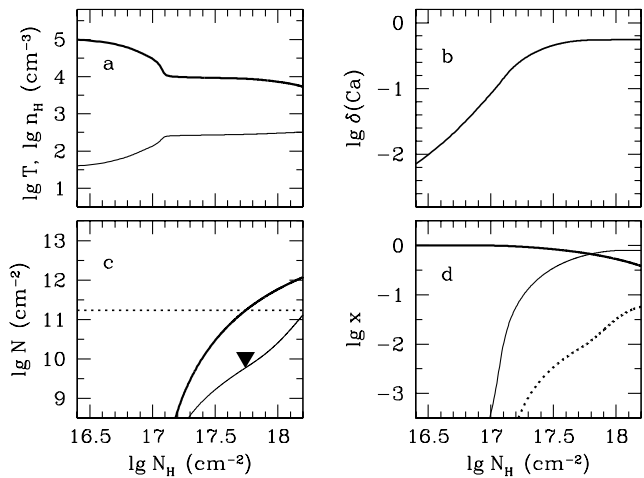}}
\caption{The same as Fig.~\ref{fig:fsh1} but for the shock speed
of 100~\kms}
\label{fig:fsh2}
\end{figure}

We focus on the approaching cloud (cloud~A) and consider two cases (I) the
cloud was shocked recently with the shock speed of 140~\kms; (II) the cloud was
shocked a long time ago by the cloud shock with the speed of 100~~\kms\ and
afterwards was accelerated up to 147~\kms. We have explored an extended
parameter space and found an optimal models which meet constraints from both
the Ca\,{\sc ii} column density and acceleration for the case~I (Fig.~\ref{fig:fsh1}).
and the case~II (Fig.~\ref{fig:fsh2}).

In case I the observed velocity 147~\kms\ is reached in $\approx 20$~yr after
the cloud crushing. The optimal value for the cloud preshock density is $n_{\rm
c0}= 4$~\cmc, while the intercloud density turns out to be $n_{\rm
i0}=0.22$~\cmc\ assuming $v_{\rm b}=600$~\kms. Shown in Fig.~\ref{fig:fsh1}
as functions of the total hydrogen column density are the temperature, 
the total hydrogen
number density, the Ca depletion parameter, the gaseous Ca\,{\sc ii} and Na\,{\sc i}
downstream column density, and the ionization fractions of H, Ca\,{\sc ii}, and Ca\,I.
The column density of Ca\,{\sc ii} is sampled primarily in the cooling zone, at about
$N_{\rm H}\sim4\times10^{17}$~\cmt. The rapid increase of the Ca\,{\sc ii} density at
this distance is an outcome of the density increase in the cooling zone which
is followed by the recombination of Ca\,{\sc iii} into Ca\,{\sc ii} and the grain
destruction. The dominant mechanism of the dust destruction is the non-thermal
sputtering due to grain collisions with He atoms and ions. We find that results
are robust with respect to the initial depletion factor. The observed Ca\,{\sc ii}
column density for the cloud~A is attained in case~I at $N_{\rm
H}\sim6.6\times10^{17}$~\cmt\ (Fig.~\ref{fig:fsh1}c). This can be considered as 
an estimate of the initial 
column density of the undisturbed cloud~A. The acceleration of the shocked 
cloud with this column density is $g=1.2\times10^{-3}$~\cms\ which coincides
within errors with the observed deprojected acceleration of the cloud~A.
For the adopted cloud column density the radius of the unshocked cloud is
$a=1.2\times10^{17}$~cm and the cloud crushing time is $t_{\rm cc}=370$~yr.
Remarkably, the model Na\,{\sc i} column density is consistent with the $3\sigma$
upper limit derived from the non-detection of the high-velocity Na\,{\sc i}~D$_2$
absorption (Fig.~\ref{fig:fsh1}c). 

In the case~II (Fig.~\ref{fig:fsh2}) the observed Ca\,{\sc ii} column density and
acceleration are satisfied for the total hydrogen column density of $N_{\rm
H}=5.5\times10^{17}$~\cmt; this value is presumably the column density of
cloud~A in case~II. The cloud number density in this model is $n_{\rm c0}=
6$~\cmc\ and the intercloud density is $n_{\rm i0}=0.16$~\cmc\  assuming $v_{\rm
b}=600$~\kms. The cloud is crushed at $t_{\rm cc}=290$~yr while the observed
velocity $v_{\rm c}=147$~\kms\ is attained at $t=1.4t_{\rm cc}\sim 400$~yr. 
Remarkably, both cases I and II suggest that the shocked cloud is observed 
at about 400 yr after the interaction began. 

The above models demonstrate that for the shock velocity in the range of
100--140~\kms\ the cloud total hydrogen column density turns out to be in a 
narrow range of
(5.5--6.6)$\times10^{17}$~\cmt. The number density in the undisturbed cloud is
in the range of 4--6~\cmc\  with the lower value corresponding to the larger shock
velocity, while the cloud radius is in the range of 
(0.7--1.2)$\times10^{17}$~\cmt\ 
with the lower limit corresponding to the lower shock velocity. 
 If the cloud shape is clumpy, we come to the estimate of 
the cloud mass in the range of $10^{-3} - 10^{-2}~M_{\odot}$. 

We also have performed simulations for the cloud~B and find that the 
Ca\,{\sc ii} column density and acceleration are reproduced in cases of 
high and low cloud shock velocity for the cloud column density in the 
range of $(4-5.4)\times10^{17}$~\cmt\ and 
cloud number density in the range of 3.8--5.5~\cmc, both values being 
similar to those for the cloud A.
The fact that both clouds lying at the opposite sides of Vela SNR have 
comparable parameters indicates that low column density clouds may be 
rather common at least for the vicinity of the Vela SNR. We emphasise 
that this conclusion is rather speculative and requires confirmation 
by means of similar study of other high-velocity components in 
Vela SNR.

\section{Discussion and Conclusions}

Our goal was to study the interstellar absorptions in spectra of stars towards
\rx\ focusing primarily on the high velocity components associated with shocked
clouds. We find that the low velocity absorption components do not show
significant anomalies when compared to properties of absorbers in other surveys.
The major differences with other surveys are probably caused by 
the selection effects due to different spectral resolution and S/N ratios.

We find an interesting effect in the distribution of K\,{\sc i} absorbers 
along the distance of background stars towards \rx: the lack of noticeable 
K\,{\sc i} absorbers at the distances $d<600$~pc followed by the "jump" of the
strength of K\,{\sc i} absorbers for $d>600$~pc which is indicative of the
underpopulation of dense clouds at $d<600$~pc. Generally, the distribution of
H\,I gas in external spiral galaxies is intermittent with characteristic
scale of several hundred parsecs, which is seen clearly, e.g., in M\,51 21~cm
map \citep{1990AJ....100..387R}.  It might well be that we look at the \rx\
through such local void with small number of cool clouds. Another possibility is
that the hollow in the distribution of K\,{\sc i} absorbing clouds might be outcome of
the Gum nebula dynamical evolution. Indeed, the Gum supershell is shown to be depopulated
by neutral gas clouds between 350--570~pc \citep{2001MNRAS.325.1213W}.

We have detected four high-velocity (|$\Vlsr$|$>100$~\kms) Ca\,{\sc ii} clouds in 
spectra of three
stars. Towards HD\,75309 we see two shocked clouds, both observed earlier in
1993 and 1996 \citep{2000ApJS..126..399C}. The large time span of 12--15~yr
permitted us to detect and measure the acceleration for cloud~A and B which turn
out to be comparable ($\sim10^{-3}$~\cms). For the first time the acceleration of
shocked clouds is used in combination with the velocity and Ca\,{\sc ii} column
density to recover the cloud total hydrogen column density
($\sim6\times10^{17}$~\cmt), which implies the cloud radius of $\sim10^{17}$~cm 
for the estimated cloud density $n_{\rm c0}\sim5$~\cmc. Regardless of 
our plane shock
model is rather crude for the description of a blast wave/cloud interaction 
we believe that our model grasps the major features of this phenomenon. 

The fact that approaching and receding high-velocity clouds on the opposite 
sides of Vela SNR
have comparable parameters indicates that they are possibly rather
typical small scale ($a\sim(1-2)\times10^{17}$ cm)
clouds in the vicinity of Vela SNR.
It is not clear whether all the high-velocity clouds observed in Vela SNR 
have similar column density but, if we admit they do, it would be 
instructive to check the outcome of this conjecture. In our sample of 14
stars eleven do not show fast components which translates into the occultation
optical depth (average number of clouds on the sight line) $\tau_{\rm oc}=0.24$.
This value is sampled at front and rear sides of the blast wave with the
thickness $\Delta\,R\approx0.03R$. Adopting the cloud number density
distribution $d\nu/da\propto\,a^{-3}$ (cf.~Appendix~\ref{appA}) which follows
from the observed probability density function of H\,I column density
$p(N)\propto\,N^{-1}$ \citep{2005ApJ...631..371S} and using the estimate
$\tau_{\rm oc}=0.24$ one finds the volume filling factor of clouds with the
radii $1<a_{17}<2$ to be $f\approx0.016$. For the cloud total hydrogen 
density $\sim
5$~\cmc\ the filling factor of the small size clouds extrapolated to larger
sizes should not exceed 0.1 in order to be consistent with the average IS
density of $\sim0.5$~\cmc.  This means that the upper limit of cloud 
size for that particular variety is $\sim4\times10^{17}$~cm. 
We therefore should admit that either the accelerated clouds are very 
rare and we catch an improbable chance, or, alternatively, the small-size clouds
are rather numerous. If this is the case then possibly this variety of 
small clouds have not yet been observed. 
Indeed, if calcium were depleted in these clouds
with $\delta=2\times10^{-4}$ then the Ca\,{\sc ii} column density 
would not exceed $2\times10^8$~\cmt\ which is below the detection limit. 
On the other hand,  if Ca depletion
were moderate, e.g., $\delta\sim2\times10^{-2}$ likewise in clouds
studied by \citet{1993A&A...278..549B} then the Ca\,{\sc ii} column density 
would be
as high as $2\times10^{10}$~\cmt\ and these clouds could fall in the range 
of observed thinnest Ca\,{\sc ii} absorbers.

At a first glance these small clouds could be identified with thin clouds
($N_{\rm H}\sim1.3\times10^{18}$~\cmt) revealed by 21~cm absorption
\citep{2005A&A...436L..53B, 2005ApJ...631..371S}. The problem, however, 
is that
these H\,I clouds are cool ($T\sim 50$~K) and therefore dense 
($n({\rm H})\sim30$ cm$^{-3}$), factor $\sim5$ more dense than our clouds.
Another problem their low filling factor. Indeed, on the sight line
towards 3C286 extragalactic source ($b=81^{\circ}$) only three clouds are 
detected, i.e., the
average number of clouds on the sight line $\tau_{\rm oc}$=3. This value is
translated into the filling factor $f$ of clouds using the relation 
\begin{equation}
\tau_{\rm oc}=\frac{3}{4}\left(\frac{h_z}{a}\right)f\,,
\label{eq-tauoc}
\end{equation}
where $h_z$ is H\,I scale height perpendicular to the Galactic plane, $a$ 
is the cloud radius. For the typical density of $\sim$30~\cmc\ 
the cloud radius is 
$a\approx4\times10^{16}$~cm. Inserting into the equation
(\ref{eq-tauoc}) the values of $a$, $\tau_{\rm oc}$, and $h_z$=100~pc we get
$f\approx5\times10^{-4}$,  which is $\approx30$ times smaller than the filling
factor of the small clouds around Vela SNR derived assuming that all 
the high-velocity 
clouds are small-size clouds. This disparity and larger density of 
thin 21 cm absorbers imply, that the two varieties of small clouds seem 
to be different.

Ultraviolet {\em HST} spectroscopy reveals in the direction
$(l,b)\sim(83^{\circ}, -50^{\circ})$ a cloud with the column density  of $N_{\rm
H}\sim6\times10^{17}$~\cmt\ and number density of 20--45~\cmc\
\citep{2007ApJ...668.1012W}. Again, this cloud has comparable column density 
with our A and B clouds but the number density is of one order magnitude 
larger. In fact, the density of A and B clouds ($\sim5$~\cmc) is 
somewhat unusual: it falls into unstable region of phase diagram for the typical
pressure $P/k\approx3000$~\cmc\,K \citep{2003ApJ...587..278W}. This means that
either A and B cloud were in the unstable state, or the pressure in the
vicinity of Vela SNR is significantly (one order of magnitude) larger 
than the typical pressure. The latter, however, seems less likely because 
according to the extensive survey of UV fine-structure C\,I absorption lines 
the typical pressure lies in the range
$10^3\lesssim P/k\lesssim10^4$ \citep{2001ApJS..137..297J}.

 To conclude, the detection of the acceleration of two shocked clouds 
in Vela SNR along the same sight line and the fact of similarity of their 
parameters are exciting results which might have interesting 
implications for our understanding of cloudy structure of the ISM at 
small scales of $\sim10^{17}$~cm. 
However, the issue, whether these clouds are common for diffuse ISM, 
or we observe highly improbable event, remains to be explored, e.g., 
by means of the study the velocity evolution of other high-velocity 
absorbing clouds in Vela SNR.

\section{Acknowledgements}

This study is partially supported by the Program of State Support to Leading 
Scientific Schools of the Russian Federation (grant 3602.2012.2) and by the
Basic Research Program of the Russian Academy of Sciences ``Nonstationary
phenomena in the Universe''.

\appendix
\section{Relation between observed and true distribution of cloud sizes}
\label{appA}

We consider homogeneous spatial distribution of spherical clouds of the same
density. A line segment of a length $L$ produced by a random sight line 
can be considered as a proxy for the corresponding column density $N$.
The probability density function of segments for a cloud of the radius $a$ is
$p(L,a)=(1/2)L/a^2$. Let the true cloud number density distribution of cloud
radii be $d\nu/da=Ca^{-\gamma}$, while $dn/dN\propto dn/dL$ is the observed
column density (or segment) distribution per unit length along some line of
sight. We then have an obvious relation between $d\nu/da$ and $dn/dL$
$$
dn/dL=\int_{L/2}^{a_{\rm max}}\pi a^2(d\nu/da)p(L,a)da \propto L^{-(\gamma-2)}\,,
$$
where we assume that $\gamma>2$ and $a_{\rm max}\gg L$. For the case 
$dn/dN\propto N^{-1}$ suggested by 21 cm observations \citep{2005ApJ...631..371S}
one gets $\gamma=3$.

\bibliographystyle{mn2e}
\bibliography{paper}

\end{document}